\documentclass[twocolumn,twoside,slac_two]{revtex4}
\usepackage{graphicx}
\usepackage{fancyhdr}
\renewcommand{\headrulewidth}{0pt}
\renewcommand{\footrulewidth}{0pt}

\newcommand{\etal}{et al.}
\newcommand{\ess}{\ensuremath{S(\mkpi)}}

\newcommand{\pee}{\ensuremath{P(\mkpi)}}

\newcommand{\dee}{\ensuremath{D(\mkpi)}}

\newcommand{\SP}{\ensuremath{2~\ess \times \pee}}

\newcommand{\PD}{\ensuremath{2~\pee \times \dee}}
\newcommand{\Dint}{\ensuremath{D^2~(\mkpi)}}
\newcommand{\Sint}{\ensuremath{S^2~(\mkpi)}}
\newcommand{\Pint}{\ensuremath{P^2~(\mkpi)}}

\newcommand{\ksw}{\ensuremath{\overline{K}_0^{*0}(1430)}}
\newcommand{\Hspls}{\ensuremath{H^2_+(\qsq)}}
\newcommand{\Hspm}{\ensuremath{H^2_\pm(\qsq)}}
\newcommand{\Hsmin}{\ensuremath{H^2_-(\qsq)}}
\newcommand{\Hszer}{\ensuremath{H^2_0(\qsq)}}
\newcommand{\Hpls}{\ensuremath{H_+(\qsq)}}
\newcommand{\Hmin}{\ensuremath{H_-(\qsq)}}
\newcommand{\Hzer}{\ensuremath{H_0(\qsq)}}
\newcommand{\hzer}{\ensuremath{h_0(\qsq)}}

\newcommand{\Hint}{\ensuremath{\hzer\,\Hzer}}

\newcommand{\klndk}{\ensuremath{D^0 \rightarrow K^- \ell^+ \nu }}

\newcommand{\krzb}{\ensuremath{\overline{K}^{*0}}}
\newcommand{\krzmndk}{\ensuremath{D^+ \rightarrow \krzb \mu^+ \nu}}
\newcommand{\krzlndk}{\ensuremath{D^+ \rightarrow \krzb \ell^+ \nu_\ell}}
\newcommand{\philndk}{\ensuremath{D_s^+ \rightarrow \phi\; \ell^+ \nu_\ell}}

\newcommand{\kkendk}{\ensuremath{D_s^+ \rightarrow K^+ K^- \ell^+ \nu }}
\newcommand{\kklndk}{\ensuremath{D_s^+ \rightarrow K^+ K^- e^+ \nu }}
\newcommand{\phiendk}{\ensuremath{D_s^+ \rightarrow \phi\; e^+ \nu_e }}
\newcommand{\kpimndk}{\ensuremath{D^+ \rightarrow K^- \pi^+ \mu^+ \nu }}
\newcommand{\kpiendk}{\ensuremath{D^+ \rightarrow K^- \pi^+ e^+ \nu }}
\newcommand{\kpilndk}{\ensuremath{D^+ \rightarrow K^- \pi^+ \ell^+ \nu }}
\newcommand{\veclndk}{\ensuremath{D \rightarrow {\rm vector}~ \ell^+ \nu }}
\newcommand{\PSlndk}{\ensuremath{D \rightarrow {\rm pseudoscalar}~ \ell^+ \nu }}

\newcommand{\thv}{\ensuremath{\theta_\textrm{v}}}
\newcommand{\thl}{\ensuremath{\theta_\ell}}
\newcommand{\costhv}{\ensuremath{\cos\thv}}

\newcommand{\costhl}{\ensuremath{\cos\thl}}

\newcommand{\sinthlsq}{\ensuremath{\sin^2\thl}}
\newcommand{\qsq}{\ensuremath{q^2}}

\newcommand{\bw}{\ensuremath{\textrm{BW}}}
\newcommand{\mkpi}{\ensuremath{m_{K\pi}}}

\newcommand{\rtwo}{\ensuremath{r_2}}

\newcommand{\rvee}{\ensuremath{r_v}}
\newcommand{\mysection}[1]{\section{#1}}

\begin{document}

\title{Recent results on 4-body, charm semileptonic decays}

%

\author{Jim Wiss}
\affiliation{University of Illinois, 1110 W. Green, Urbana IL , 61801}

\begin{abstract}
We summarize recent data on 4-body charm semileptonic decay concentrating on \kklndk{} and \kpilndk{}.
We begin with giving some motivation for the study of these decays. We discuss several of 
the models traditionally used to describe these decays and conclude by presenting a non-parametric
analysis of \kpiendk{} and its possible extension into non-parametric studies of \kpimndk{}.
\end{abstract}
\maketitle

\thispagestyle{fancy}


\section{Introduction}

\begin{figure}[tbph!]
 \begin{center}
  \includegraphics[width=3.5in]{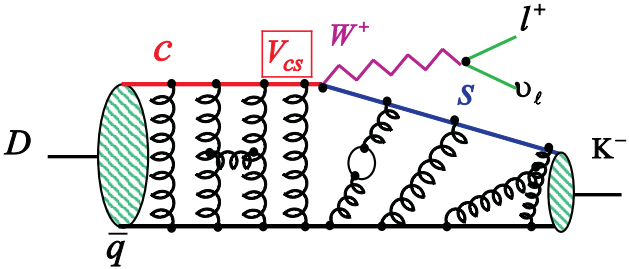}
  \caption{Diagrams for the semileptonic decay of charmed mesons. The
hadronic,QCD complications are contained in \qsq{} dependent form factors. 
\label{cartoon}} \end{center}
\end{figure}

Figure \ref{cartoon} shows a cartoon of the $D^0 \rightarrow K^- \ell^+ \nu$ decay process.  
All of the  hadronic complications for this process is contained
in $\qsq{}$ dependent form factors that are computable 
using non-perturbative methods such as LQCD.  Although semi-leptonic process can in principle provide a  determination of charm CKM elements, one frequently uses the (unitarity constrained) CKM measurements, lifetime, and branching
fraction to measure the scale of charm semileptonic decay constants and compare them to 
LQCD predictions. The $\qsq{}$ dependence of the semileptonic form factor can also be directly measured
and compared to theoretical predictions.  

The hope is that charm semileptonic decays can provide high statistics, precise tests of LQCD calculations and thus
validate the computational techniques for charm.  
Once validated, the same LQCD techniques can be used in related calculations for $B$-decay
and thus produce CKM parameters with significantly reduced theory systematics.  

Although recent, unquenched LQCD calculations are unavailable for \veclndk{} processes, owing to the instability 
of the vector parent, I hope that the 4-body will provide additional tests of LQCD for a variety of spin
states which will further help calibrate the lattice, and provide confidence in analogous decays for the beauty sector.

\begin{figure}[tbph!]
 \begin{center}
  \includegraphics[width=2.in]{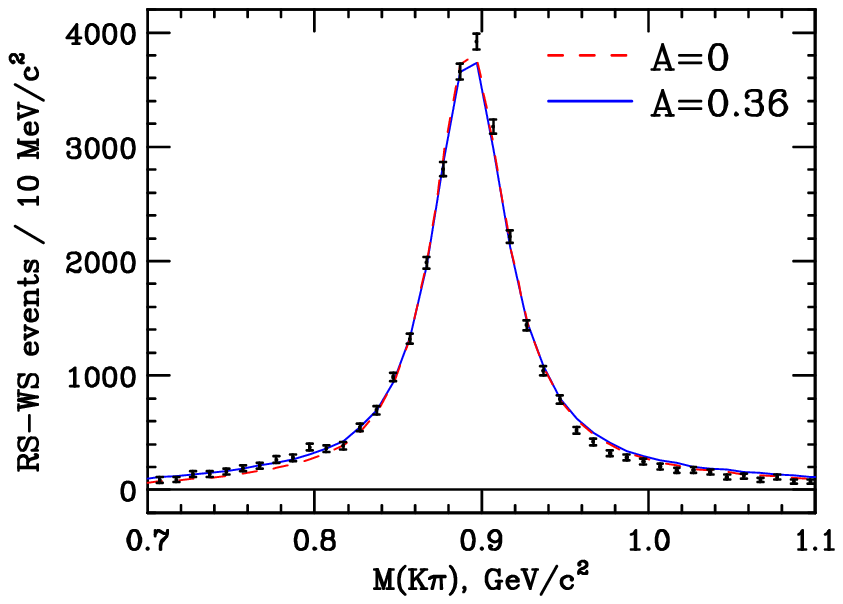}
  \includegraphics[width=2.in]{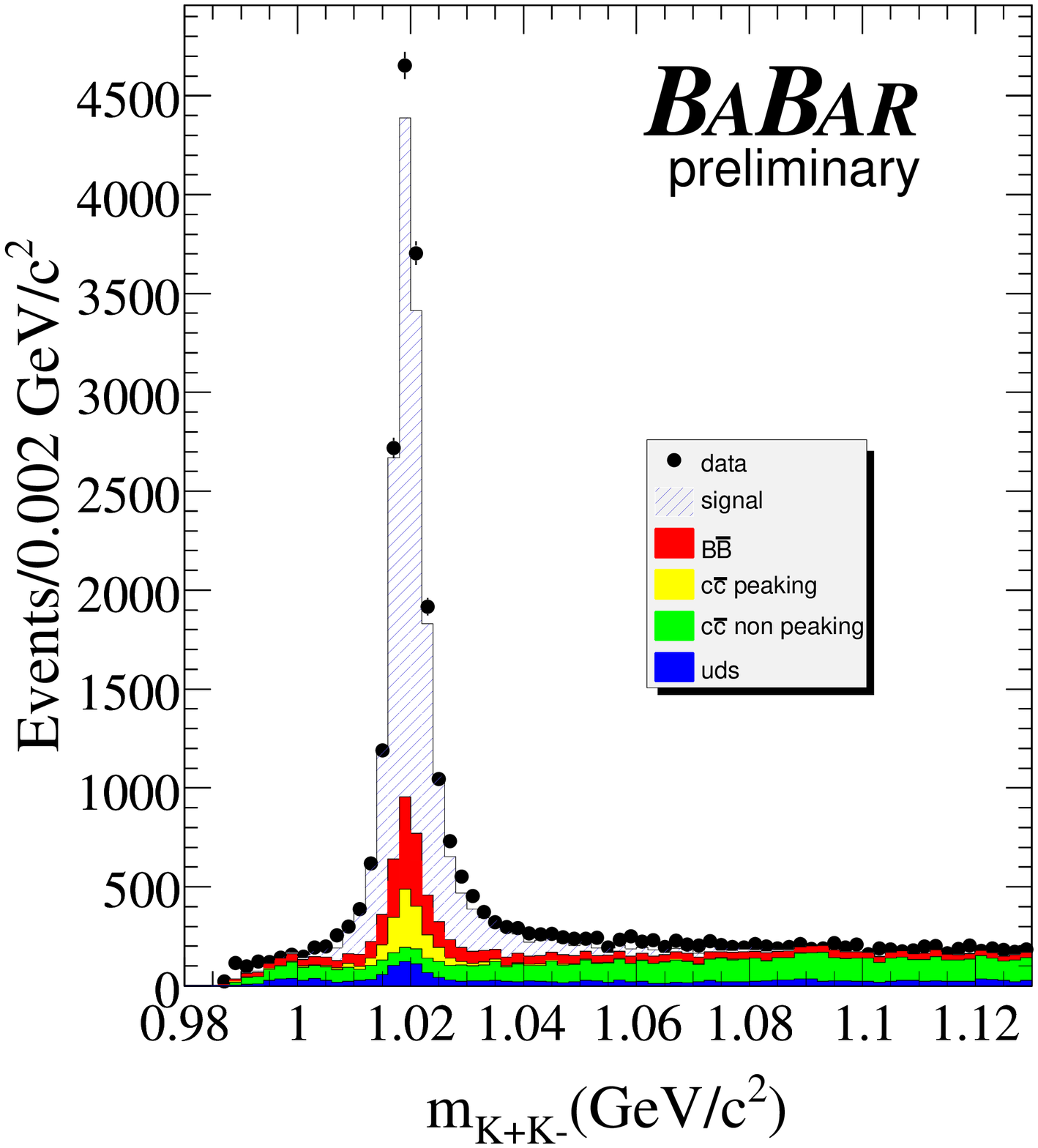}
  \caption{   We show the $m(K^+K^-)$ spectra obtained in \kkendk{} by BaBar\cite{BBPhilnu} and 
$m(K^-\pi^+)$ by FOCUS\cite{swave}. The curve on the FOCUS $m(K^-\pi^+)$ spectra is a \krzb{} line shape
both with (A = 0.36) and without (A = 0) a small s-wave, non-resonant component which was found through an interference
in the decay intensity and is described later. The $m(K^+K^-)$ spectra obtained by BaBar is very strongly dominated by the $\phi$
resonance along with a few known backgrounds. 
\label{VecDom}} \end{center}
\end{figure}

I find it remarkable that 4-body semileptonic decays such as \kkendk{} and \kpimndk{} are so heavily
dominated by the vector decays \phiendk{} and \krzmndk{}. Figure \ref{VecDom} illustrates this dominance by showing
data from FOCUS\cite{swave} and recent data from BaBar\cite{BBPhilnu}.
The absence of a substantial non-resonant, or higher spin resonance component to these decays means
the decay angular distribution can be described in terms of three, \qsq{}-dependent helicity
basis form factors that describe the coupling of the lepton system to the three helicity states
of the vector meson according to Eq. (\ref{dkelect}) :

\begin{eqnarray}
\left| \cal{A} \right|^2  \approx \frac{{q^2 }}{8}\left| \begin{array}{l}
 (1 + \cos \theta _l )\sin \theta _V e^{i\chi } H_ + (\qsq) \\ 
  - (1 - \cos \theta _l )\sin \theta _V e^{ - i\chi } H_ - (\qsq)  \\ 
  - 2\sin \theta _l \cos \theta _V H_0 (\qsq) \\ 
 \end{array} \right|^2
\label{dkelect}
\end{eqnarray}

\begin{figure}[tbph!]
 \begin{center}
  \includegraphics[width=2.in]{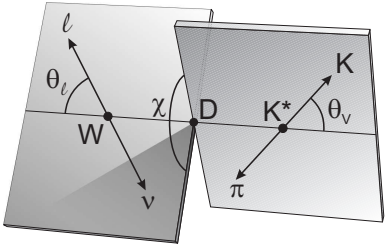}
    \caption{Definition of kinematic variables.
 \label{anglesA}}
 \end{center}
\end{figure}

The three decay angles describing the \kpilndk{}
decay, referenced in Eq.(\ref{dkelect}), are illustrated by Fig. \ref{anglesA}. 
\mysection{Analytic models for form factors}

We begin by describing the three form factors relevant to \veclndk{} although there is strong evidence \cite{swave}\cite{cleo-ff} for a non-resonant, s-wave component to \kpilndk{}. A new, fifth form factor $H_T(\qsq)$ is also required for \krzmndk{} to describe
the suppressed coupling of the \krzb{} to a left-handed $\mu^+$.

The   \Hpls{} , \Hmin{}, \Hzer{} form factors are linear combinations of two axial and one vector form factor~\cite{KS} according to Eq. ~(\ref{helicity}): 
\begin{widetext}
\begin{eqnarray}
H_\pm(\qsq) &=&
   (M_D+\mkpi)A_1(\qsq)\mp 2{M_D K\over M_D+m_{K\pi}}V(\qsq) \,,
                               \nonumber \\
H_0(\qsq) &=&
   {1\over 2\mkpi\sqrt{\qsq}}
   \left[
    (M^2_D -m^2_{K\pi}-\qsq)(M_D+\mkpi)A_1(\qsq) 
    -4{M^2_D K^2\over M_D+\mkpi}A_2(\qsq) \right] \label{helicity} \nonumber \\ 
\label{helform}
\end{eqnarray}
\end{widetext}
where $K$ is the momentum of the $K^- \pi^+$ system and \mkpi{} is its mass.

\begin{figure}[tbph!]
\begin{center}
  \includegraphics[width=2.5in]{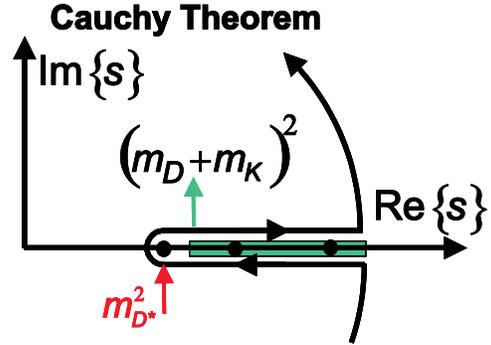}
  \caption{ Each form factor is assumed to be an analytic function with pole singularities at the 
masses of bound states, and cuts that start at the start of the continuum. We illustrate
the case of \klndk{}. One can use Cauchy's
theorem with the indicated contour to write an dispersion expression for each form factor in the physical range 
$0 < \qsq{} < \left(m_D - m_K \right)^2$ 
\label{cut}} 
\end{center}
\end{figure}

Eq.(\ref{analytic}) provides considerable insight into the expected analytic form for semileptonic form factors.
It uses a dispersion relation obtained using Cauchy's Theorem under the assumption that a form factor is an analytic, complex function apart from some known singularities. Fig. \ref{cut} illustrates the Cauchy's Theorem contour for the case for the $f_+(\qsq{})$ form factor
describing \klndk{}. 

The form factor singularities will consist of a sum of simple poles at the D meson -kaon vector bound states (e.g. $D_s^{*+}$) plus
a cut beginning at the $D-{\rm kaon}$ continuum in the cross process: $\nu \ell^+ \rightarrow D~{\rm kaon}$. The dispersion relation
gives the form factor ($\rm{F}(\qsq)$) as a sum over the spectroscopic poles plus an integral over the cut.
\begin{eqnarray}
\rm{F}(q^2 ) = \frac{\mathcal{R}}{{m_{D_s^*}^2  - q^2 }} + \frac{1}{\pi }\int_{\left( {m_D  + K} 
\right)^2 }^\infty  {\frac{{{\mathop{\rm Im}\nolimits} \left\{ {f_ +  (s)} \right\}}}{{s - q^2  - 
i\varepsilon }}ds} 
\label{analytic}
\end{eqnarray}

Both the cuts and poles are generally beyond the physical $\qsq{}_{\rm max}$ and thus can never be actually realized. 

Spectroscopic pole dominance (SPD) was an early parameterization for the form factors relevant to both \veclndk{} and \PSlndk{}.
SPD ignores the cut integral entirely and approximates $\rm{F}(q^2 )$ using just the first term of Eq.(\ref{analytic}). 
The advantage of SPD approach is that it requires only a single unknown fitting parameter $\mathcal{R}$ to describe  
each $\rm{F}(\qsq)$ since the positions of the bound states are well known. SPD entirely predicts {\it shape} of \PSlndk{} decay intensity
and predicts that the {\it shape} for the \krzlndk{} can be fit by just two parameters
which are traditionally taken to be the axial and vector form factor ratios at $\qsq = 0$: $\rvee = V(0)/A_1(0)$ and $\rtwo = A_2(0)/A_1(0)$.

\begin{figure}[tbph!]
 \begin{center}
  \includegraphics[width=2.75in]{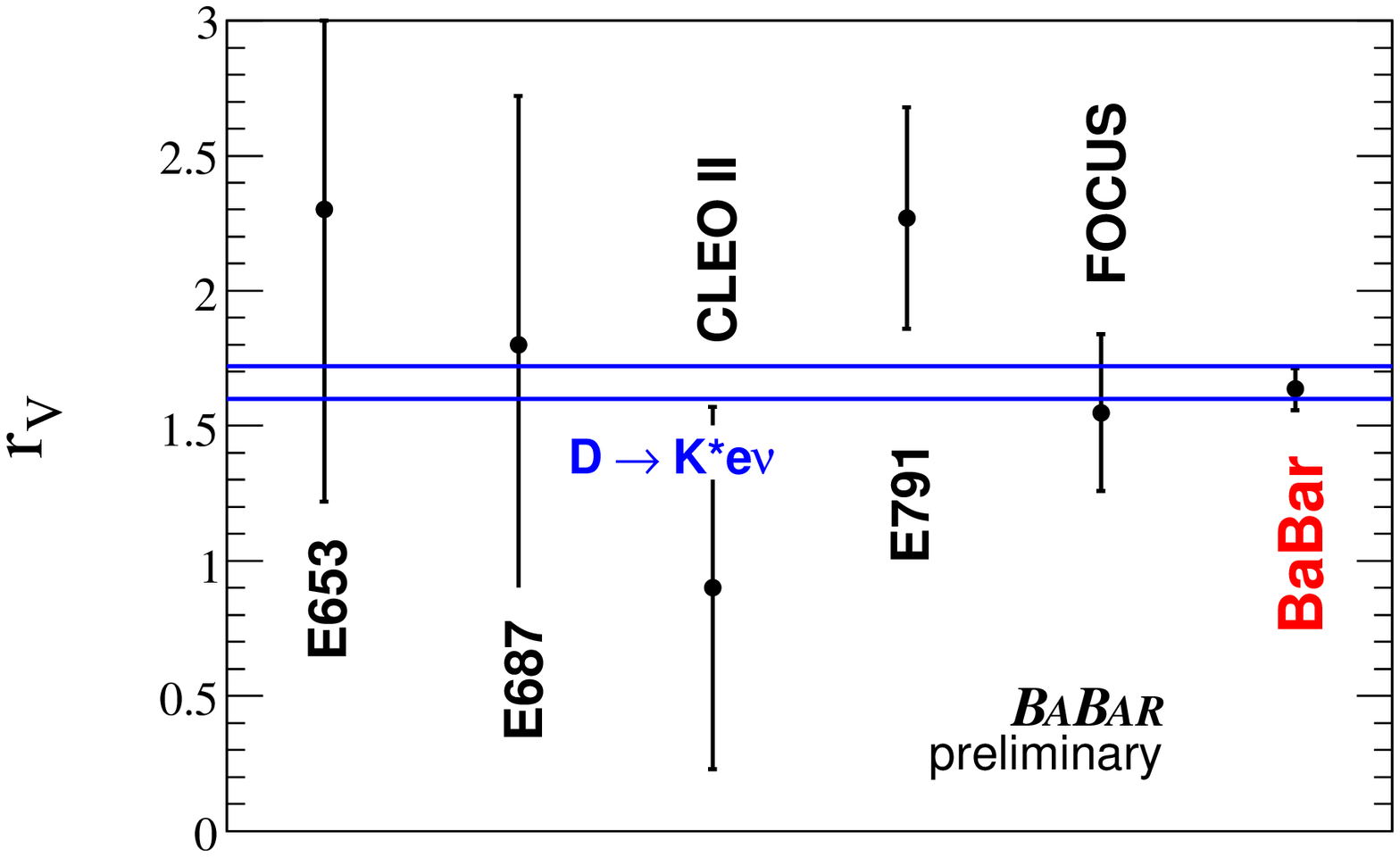}
\includegraphics[width=2.75in]{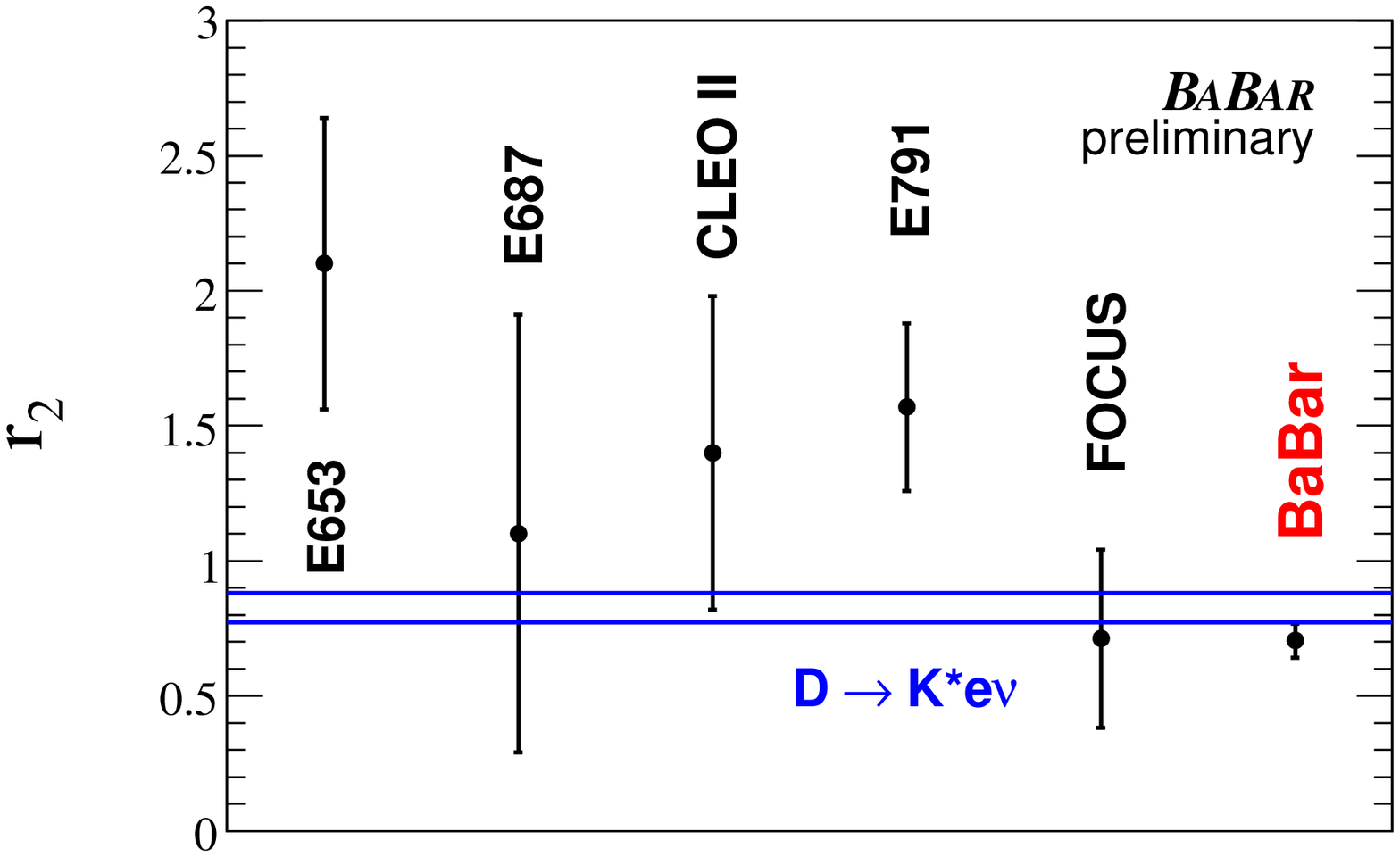}
  \caption{   
The \rvee{} and \rtwo{} form factor ratios measured for \philndk{} by various experiments. The blue
lines show $\pm 1 \sigma$ bands for the weighted average of the \krzlndk{} form factor ratios compiled in Reference \cite{fpcp}. It is
expected from SU(3) symmetry that the \philndk{} form factors should be very close to those for
\krzlndk{}
 \label{phi-ff}} \end{center}
\end{figure}

BaBar \cite{BBPhilnu} has recently published an interesting SU(3) test based on SPD applied to \phiendk{}. 
Figure \ref{phi-ff} compares the \rvee{} and \rtwo{} parameters measured for \philndk{} to those previously measured for
\krzlndk{}. By SU(3) symmetry and explicit calculation, the \rvee{} and \rtwo{} form factor ratios for \krzlndk{} and \philndk{} decays are
expected to be very close to each other. This is true for \rvee{}, but previous to the recent measurement by the FOCUS Collaboration\cite{focusPhi}, \rtwo{} for  \philndk{} was measured to be roughly a factor of two larger than that
for \krzlndk. BaBar\cite{BBPhilnu} has confirmed the expected consistency between the form factor ratios obtained for 
\philndk{} and \krzlndk{} with unparalleled statistics.

\begin{figure}[tbph!]
 \begin{center}
  \includegraphics[width=3.3in]{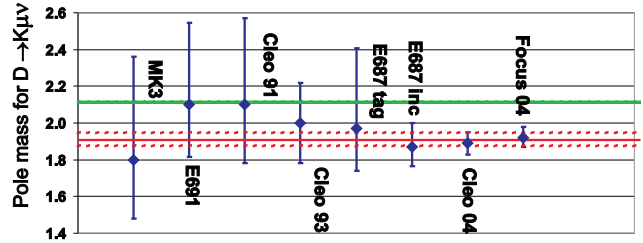}
  \caption{ Effective pole mass measurement in $D^0 \rightarrow K^- e^+ \nu$ over the years.
The green line is the $m_{D_S^*}$ sectroscopic pole mass and is inconsistent with the 
average of the displayed data by 5.1 $\sigma$. \label{mpole}} \end{center}
\end{figure}

Several experiments have tested SPD by measuring an ``effective" pole mass ($m_{pole}$)in $D^0 \rightarrow K^- e^+ \nu$ decay 
where the pole mass is defined using ${\rm{f}}_{\rm{ + }} (q^2 ) \propto 1/{(m_{pole}^2  - q^2)}$.  As Fig. \ref{mpole} from Reference \cite{fpcp} shows, as errors have
improved over the years, it becomes clear that effective pole is significantly lower than the spectroscopic 
pole, underscoring the importance of the cut integral contribution for this decay.

\begin{figure}[tbph!]
 \begin{center}
  \includegraphics[width=3.2in]{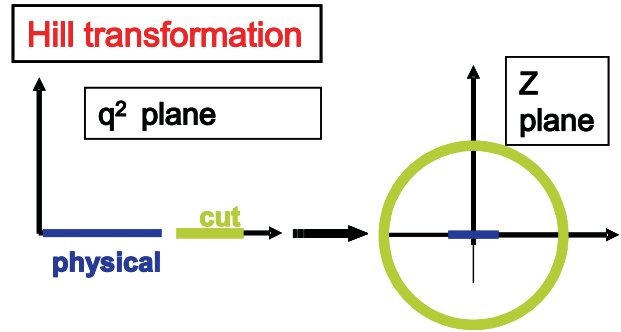}
  \caption{Illustration of Hill transformation approach. \label{hillTrans}} \end{center} \end{figure}

Several parameterizations have been proposed to include the cut integral in Eq. (\ref{analytic}) as well as the spectroscopic poles. 
Becirevic and Kaidalov (1999) \cite{BK} proposed a new parameterization for the \PSlndk{} for factor $f_+(\qsq{})$ that replaces the cut integral by an effective pole where the heavy quark symmetry and other theoretical ideas are used to relate the residue and effective pole position. These constraints
leads to a modified pole form with a single additional parameter $\alpha$ that describes the degree to which
the single spectroscopic pole fails to match $f_+(\qsq{})$ for a given process. 

\begin{eqnarray}
f_ +  (q^2 ) = \frac{{f_ +  (0)}}{{\left( {1 - q^2 /m_{D*}^2 } \right)\left( {1 - \alpha q^2 /m_{D*}^2 } \right)}}
\label{modpole}
\end{eqnarray}
S. Fajfer and J. Kamenik \cite{FK} have recently extended the effective pole approach to the three helicity form 
factors relevant to \veclndk{} decays.

R.J. Hill\cite{Hill2}\cite{Hill} has proposed an alternative way of viewing form factors which is illustrated in Fig. \ref{hillTrans}.
The basic idea is to devise a transformation of a form factor from the complex \qsq{} plane to a complex $z$ plane. This transformation is devised to (1) remove the spectroscopic poles and (2) put the cuts far away from the physical $z$ region. After the 
transformation, since the singularities
have been removed or diminished, each form factor can be well represented by a low order Taylor series in $z$.  The transformation approach is known\cite{Hill} to work very well in $B$-decays where the physical \qsq{} region gets very close to the singularities for pseudo-scalar $B$ semileptonic decay. It also works well for pseudoscalar charm pseudoscalar semileptonic decay\cite{Hill2}. 
\mysection{ \kpilndk{} Decays}
Although historically \krzlndk{} have been the most accessible semileptonic
decays in fixed target experiments owing to their ease of isolating a signal, they are significantly more complicated
to analyze than \PSlndk{}. One problem is that a separate helicity form factor is required
for each of the three helicity states of vector meson. The \qsq{} dependence of these form factors cannot
be simply measured from the \qsq{} dependence of the decay rate as is the case in \PSlndk{} but rather
must be entangled from the \qsq{} dependence of the angular distribution such as that given by Eq. (\ref{dkelect}).

\begin{figure}[tbph!]
 \begin{center}
  \includegraphics[width=3.25in]{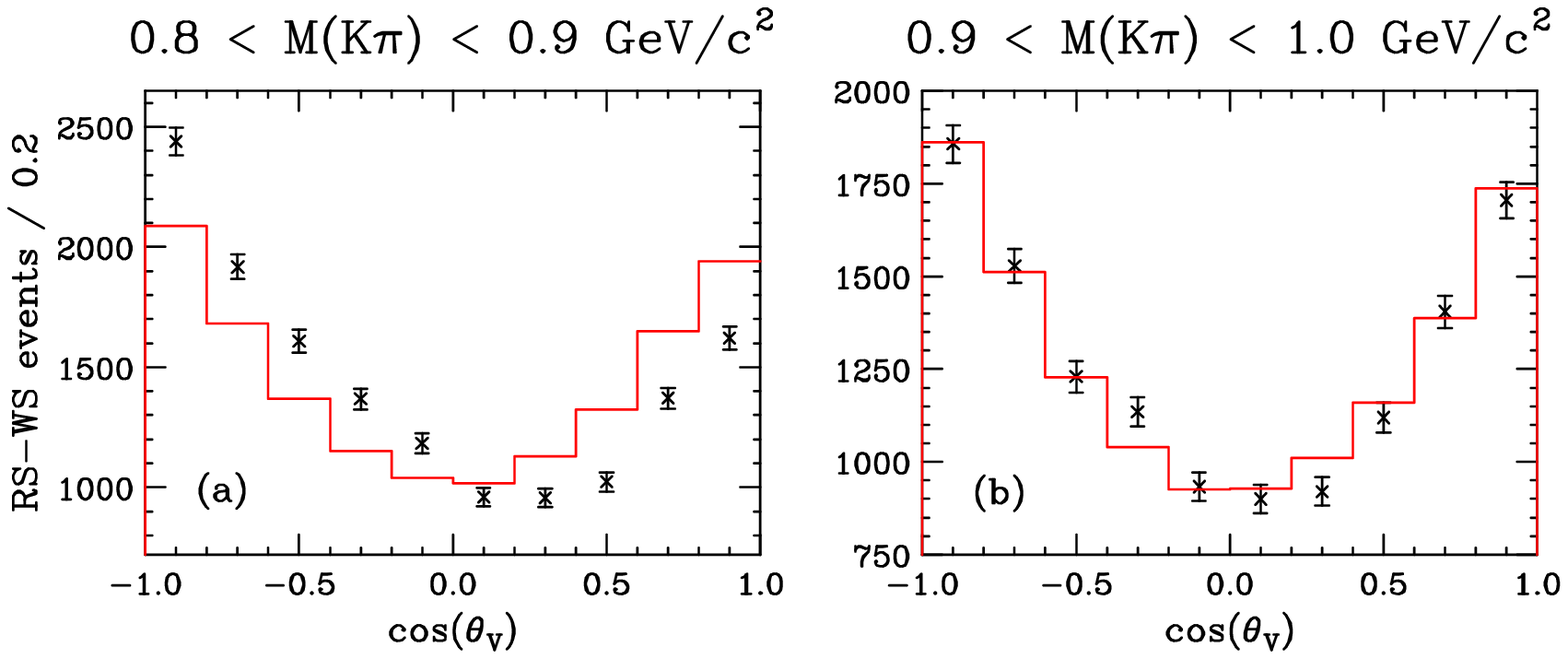}
  \caption{Evidence for s-wave interference in \kpilndk{}.
 \label{asym}}
 \end{center}
\end{figure}

Another complication is that since \kpilndk{} states result in a multihadronic
final state, the \krzlndk{} final states can potentially interfere with \kpilndk{} processes
with the $K^- \pi^+$ in various angular momentum waves with each wave requiring its own form factor.
Because the \mkpi{} distribution in \kpilndk{} was an excellent fit to the \krzb{} Breit-Wigner as shown in Fig. \ref{VecDom},
it was assumed for many years that any non-resonant component to \kpilndk{} must be negligible.
In 2002, FOCUS observed a strong, forward-backward asymmetry in \costhv{} for events with \mkpi
below the \krzb{} pole with essentially no asymmetry above the pole as shown in Figure \ref{asym}.
The simplest explanation for this asymmetry is the presence of a linear \costhv{} term in the decay intensity due to interference
between the \krzmndk{} and a non-resonant, s-wave amplitude. This interference is the second-to-last term in  Eq. (\ref{KSE}), which is basically an expanded
out version of Eq. (\ref{dkelect}), integrated over acoplanarity $\chi$. We also explicitly include the \krzb{} Breit-Wigner amplitude ($BW$). 
Note that all other interference terms (such as a possible $H_+(\qsq) \times H_-(\qsq)$ contribution) vanish because of the 
$\int_0^{2\pi} d \chi \exp(i \Delta \chi)$ integration.  Only ``same" helicity contributions can interfer in the acoplanarity averaged intensity.  We will argue shortly that an appropriate $\delta$ can create the asymmetry pattern shown in Fig. \ref{asym}

Finally we introduce an additional form factor  (\hzer{}) in
Eq. (\ref{KSE}) to describe the coupling to the s-wave amplitude.  
\begin{widetext}
\begin{eqnarray}
\int {\left| {\rm{A}} \right|^2 d\chi }  = \frac{1}{8}q^2 \left\{ \begin{array}{l}
 \left( {(1 + \cos \theta _l )\sin \theta _V } \right)^2 \left| {H_ +  (q^2 )} \right|^2 \left| {BW} \right|^2  \\ 
  + \left( {(1 - \cos \theta _l )\sin \theta _V } \right)^2 \left| {H_ -  (q^2 )} \right|^2 \left| {BW} \right|^2  \\ 
  + \left( {2\sin \theta _l \cos \theta _V } \right)^2 \left| {H_0 (q^2 )} \right|^2 \left| {BW} \right|^2  \\ 
  + 8\left( {\sin ^2 \theta _l \cos \theta _V } \right)H_0 (q^2 )h_o (q^2 ){\mathop{\rm Re}\nolimits} \left\{ {Ae^{ - i\delta } BW} \right\} \\ 
  + O(A^2 ) \\ 
 \end{array} \right\}
\label{KSE}
\end{eqnarray}
\end{widetext}
\subsection{\label{asymtopia} Asymtotic Forms}
Assuming that $A_{1,2}(\qsq)$ and $V(\qsq)$ approach a constant in the low \qsq{} limit, as expected in 
spectroscopic pole dominance, Eq. (\ref{helform}) shows $\qsq{} \rightarrow 0$, both \Hpls{} and \Hmin{} approach a constant as well. 
By way of constrast, \Hzer{} will diverge in the low \qsq{} limit according to Eq. (\ref{helform}) owing to the $1/\sqrt{\qsq}$ 
prefactor. 
Since the helicity intensity contributions are proportional to $\qsq{} H^2_\pm(\qsq{})$, according to Eq.(\ref{KSE}),
the $H_\pm$ intensity contributions vanish in this limit, while $\qsq{} H^2_0(\qsq{})$ will approach a constant.

\begin{figure}[tbph!]
 \begin{center}
  \includegraphics[width=3.0in]{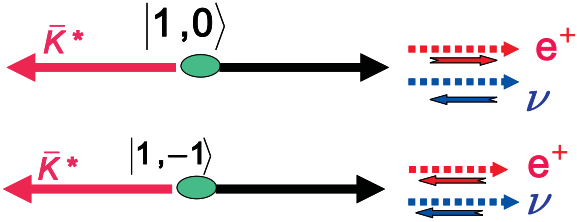}
  \caption{The electron helicity state in the low \qsq{} limit. When the virtual $W^+$ is in the zero
helicity state, the $e^+$ and $\nu$ have the opposite helicity and can be in their charged-current
helicity states. When the virtual $W^+$ is in the $| 1 , \pm 1 \rangle$ state the
$e^+$ and $\nu$ must be in the same helicity states and violate the weak helicity rules. 
 \label{hcartoon}}
 \end{center}
\end{figure}

Figure \ref{hcartoon}
expains why this is true. As $\qsq{} \rightarrow 0$, the $e^+$ and $\nu$ become collinear with the virtual $W^+$.
For \Hpls{} and \Hmin{}, the virtual $W^+$ must be in the $| 1 , \pm 1 \rangle$ state which means
that the  $e^+$ and $\nu$ must both appear as either right-handed or left-handed thus violating the charged current helicity rules. 
Hence $\qsq{}H_\pm (\qsq{})$ vanishes at low \qsq{}.  For \Hzer{}, the $W^+$ is in $| 1 , 0 \rangle$ state thus allowing the 
$e^+$ and $\nu$ to be in their (opposite) natural helicity state.  Hence at low \qsq{},  $\qsq{}\Hzer{} \rightarrow {\rm constant}$
which allows for  \krzmndk{} decays as $\qsq{}\rightarrow 0$. Presumably $\hzer{} \rightarrow 1/\sqrt{\qsq{}}$ as well since
it also describes a process with $W^+$ in the $| 1 , 0 \rangle$ state 
  
Here is a final observation on the expected asymtotic behavior of the helicity form factors. As $\qsq{} \rightarrow q^2_{\rm max}$, the
momenta of the virtual $W^+$ and \krzb{} approaches zero and \thv{} and \thl{} can no longer be defined.  This means the \krzlndk{} decay
must be isotropic and Eq. (\ref{KSE}) implies that $|H_\pm|^2 \rightarrow |H_0|^2$ as $\qsq{} \rightarrow q^2_{\rm max}$. 
A spectroscopic pole dominance model for the axial and vector form factors will automatically satisfies these asymtotic limits
according to Eq.(\ref{helform}). 

\mysection{Projection weighting technique} 

We next describe the projective weighting technique that we use to extract
the helicity basis form factors.  This technique was initially developed by the FOCUS Collaboration\cite{focus-helicity}
and applied to CLEO\cite{cleo-ff} data.  As shown in Eq. (\ref{KSE}), after integrating over acoplanarity, the decay intensity
is just a sum over four terms that consist of a form factor product times a characteristic angular distribution in \thv{} and 
\thl{}. The acoplanarity integration has significantly simplified the problem by eliminating the five of the possible six interference terms between the four form factor amplitudes with different helicities.
We begin by making a binned version of Eq. (\ref{KSE}) given by Eq. (\ref{series}), where for simplicity we only write three of the terms.

\begin{equation}
  \vec D_i = f_+(q_i^2)\:\vec m_+ + f_-(q_i^2)\:\vec m_- 
            + f_0(q_i^2)\:\vec m_0
\label{series}
\end{equation}

We use 25 joint  $\Delta \costhv  \times \Delta \costhl$ angular bins:  
5 evenly spaced bins in \costhv{} times 5 bins in \costhl{} and 6 bins in \qsq{} ($i = 0 \rightarrow 6$).
The number of \kpilndk{} events observed in each of the 25 angular bins is packed into a twenty-five component $\vec D_i$ ``data" vector.

The $f_\pm(q_i^2)$ and $f_0(q_i^2)$ are proportional to \Hspm{}, \Hszer{} averaged over the $q^2_i$ bin along with all phase space and efficiency factors. The $\vec m_\pm$ and $\vec m_0$ are the angular distributions due to each individual form factor
product packed into a 25-vector for each of the six \qsq{} bins. The acceptance and phase space corrected $m$-vectors
are obtained directly from a Monte Carlo simulation where a given form factor product is turned on and all others are turned off.
We can write Eq.~(\ref{series}) as the ``component equation" shown in  
Eq.~(\ref{system}) by forming the dot product with each of the three $m$-vectors:
\begin{widetext}
\begin{equation}
\left( {\begin{array}{*{20}c}
   {\vec m_+ \cdot \vec D_i }  \\
   {\vec m_- \cdot \vec D_i }  \\
   {\vec m_0 \cdot \vec D_i }  \\
\end{array}} \right) = \left( {\begin{array}{*{20}c}
   {\vec m_+ \cdot \vec m_+} & {\vec m_+ \cdot \vec m_-} 
      & {\vec m_+ \cdot \vec m_0 }  \\
   {\vec m_- \cdot \vec m_+} & {\vec m_- \cdot \vec m_-} 
      & {\vec m_- \cdot \vec m_0 }  \\
   {\vec m_0 \cdot \vec m_+} & {\vec m_0 \cdot \vec m_-} 
      & {\vec m_0 \cdot \vec m_0 }  \\
\end{array}} \right) \left( {\begin{array}{*{20}c}
   {f_+(q_i^2)}  \\
   {f_-(q_i^2)}  \\
   {f_0(q_i^2)}  \\
\end{array}} \right)
\label{system}
\end{equation}
The solution to Eq.~(\ref{system}) can be written as:

\begin{equation}
f_+ (q_i^2) = {}^i\vec P_+ \cdot \vec D_i\:,\
f_- (q_i^2) = {}^i\vec P_- \cdot \vec D_i\:,\
f_0 (q_i^2) = {}^i\vec P_0 \cdot \vec D_i
\label{solution}
\end{equation}
where ${}^i\vec P_\alpha$ vectors are given by Eq.~(\ref{projectors}).

\begin{equation}
\left( {\begin{array}{*{20}c}
   {{}^i\vec P_+}  \\
   {{}^i\vec P_-}  \\
   {{}^i\vec P_0}  \\
\end{array}} \right) = \left( {\begin{array}{*{20}c}
   {\vec m_+ \cdot \vec m_+} & {\vec m_+ \cdot \vec m_-}
       & {\vec m_+ \cdot \vec m_0}  \\
   {\vec m_- \cdot \vec m_+} & {\vec m_- \cdot \vec m_-}
       & {\vec m_- \cdot \vec m_0}  \\
   {\vec m_0 \cdot \vec m_+} & {\vec m_0 \cdot \vec m_-}
       & {\vec m_0 \cdot \vec m_0}  \\
\end{array}} \right)^{- 1} \left( {\begin{array}{*{20}c}
   {\vec m_+}  \\
   {\vec m_-}  \\
   {\vec m_0}  \\
\end{array}} \right)
\label{projectors}
\end{equation}
\end{widetext}
It is useful to think of forming the dot products in Eq. (\ref{solution})
by making a weighted histogram:
\begin{equation}
\vec P_+ \cdot \vec D = 
    \left[\vec P_+\right]_1 n_1 + \left[\vec P_+\right]_2 n_2 
       + \cdots \left[\vec P_+\right]_{25} n_{25}  
\label{likeweighting}
\end{equation}
Eq.~(\ref{likeweighting}) demonstrates the product $\vec P_ + \cdot \vec D$ is
equivalent to weighting the $n_1$ events in angular bin 1 by
$\left[\vec P_+\right]_1$, weighting the $n_2$ events in angular bin 2 by
$\left[\vec P_+\right]_2$, etc.
Hence each form factor product such as $f_+(q^2_i)$ can be obtained by simply weighting
the data by $\left[\vec P_+\right]_i$ where $i$ is the angular bin of the given datum.
The acceptance and phase space factors can be easily included the projective weights as well in order to directly
produce each form factor product.
Hence the (arbitrarily normalized)
form factor products $H_+^2(\qsq)$, $H_-^2(\qsq)$, and $H_0^2(\qsq)$ can then be obtained by making three weighted histograms using the efficiency rescaled ${{}^i\vec P}_+$, 
${{}^i\vec P}_-$, and ${{}^i\vec P}_0$ weights respectively.

\begin{figure}[htp]
\begin{center}
\includegraphics[width=3.2in]{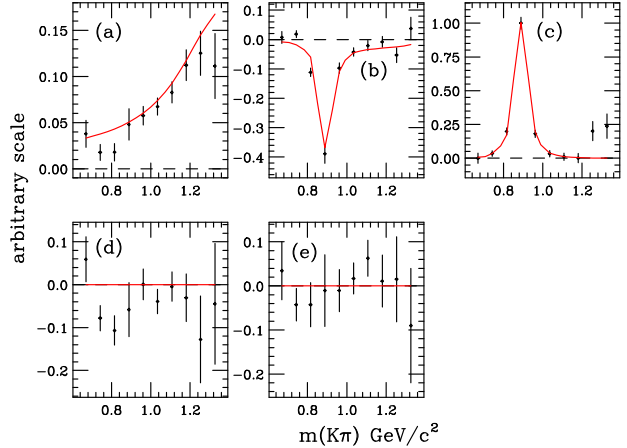}
\end{center}
\caption{ 
Results of a non-parametric analysis\cite{kkpi} of $D^+ \rightarrow K^- K^+ \pi^+$ using a variant of the projective
weighting technique described here.
The plots are:
(a)~\Sint{} direct term,
(b)~\SP{} interference term,
(c)~\Pint{} direct term,
(d)~\PD{} interference term and
(e)~\Dint{} direct term.
The overlay is a model including the \krzb{} which dominates \Pint{}, and a wider
\ksw{} which dominates \Sint{}.
\label{sysall2}}
\end{figure}

The same, basic projective weighting approach has been recently applied by the FOCUS Collaboration\cite{kkpi} for a non-parametric
analysis of the $K^- \pi^+$ amplitudes in the hadronic decay $D^+ \rightarrow K^- K^+ \pi^+$. To whet the appetite,
Fig.\ref{sysall2} shows the $K^- \pi+$ amplitudes obtained in that analysis. The s-wave amplitude shown 
in Fig. \ref{sysall2} (a) and
begs comparison with the s-wave amplitude obtained in a K-matrix analysis\cite{kpipi} of  $D^+ \rightarrow K^- \pi^+ \pi^+$ described by S. Malvezzi in these proceedings.

\section{\label{results} A non-parametric analysis of the helicity form factors in \kpilndk{}}

\begin{figure}[tbph!]
 \begin{center}
  \includegraphics[width=3.in]{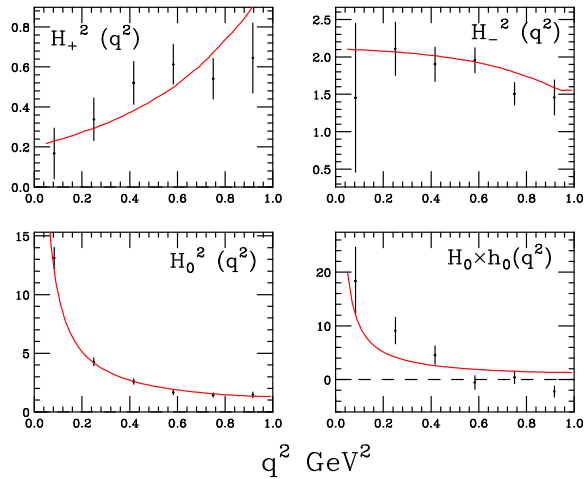}
  \caption{The four helicity form factor products obtained using the 281 $pb^{-1}$ data set
from CLEO\cite{cleo-ff}.  The curves represent the model of Reference \cite{formfactor}.
 \label{hsq}}
 \end{center}
\end{figure}

Figure \ref{hsq} shows the four weighted histograms from an analysis of 281 $pb^{-1}$ $\psi(3770)$ 
CLEO data\cite{cleo-ff}.   
Figure \ref{hsq} shows the expected behavior discussed in Section \ref{asymtopia}. In particular, $H_\pm {\qsq} \rightarrow {\rm constant}$ as $\qsq{} \rightarrow 0$ while the zero-helicity form factors, \Hzer{} and \hzer{}, diverge as $1/\sqrt{\qsq{}}$.  It is interesting to note that although the non-resonant, s-wave amplitude is too small to see in 
the $K^- \pi^+$ mass spectrum (Fig. \ref{VecDom}), its form factor is measured with roughly the same precision
as \Hspls{} or \Hsmin{}. The curves give the
helicity form factors according to Eq. (\ref{KSE}) , using spectrocopic pole dominance and 
the \rvee{}, \rtwo{}, and s-wave parameters measured by FOCUS\cite{formfactor}. Apart from the \Hint{} interference form 
factor product, the spectroscopic pole dominance model is a fairly good match to the 
CLEO non-parametric analysis.  This suggests that the ad-hoc assumption, used by FOCUS, that \hzer{}=\Hzer{} 
is questionable but it will probably take more data, and some theoretical guidance, to gain insight into the nature of the 
discrepancy.

\begin{figure}[tbph!]
 \begin{center}
  \includegraphics[width=3.in]{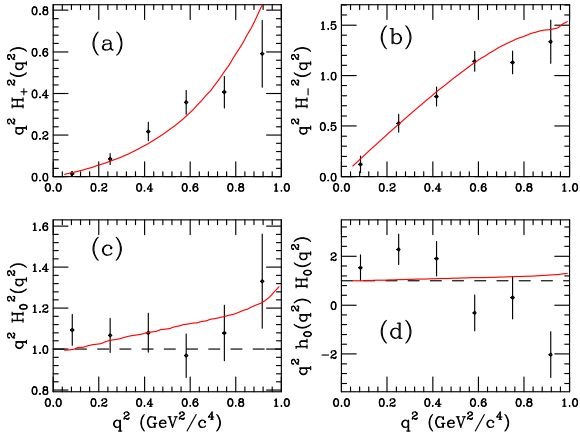}
  \caption{Non-parametric form factor products obtained for the data sample
(multiplied by \qsq{}) The reconstructed form factor products are shown as the points with error bars,
where the error bars represent the statistical uncertainties.
The solid curves in the histograms represent a form factor model described
in Ref.~\cite{formfactor}. 
The histogram plots are:
(a)~$\qsq{} H_+^2(\qsq)$,
(b)~$\qsq{} H_-^2(\qsq)$,
(c)~$\qsq{} H_0^2(\qsq)$, and
(d)~$\qsq{} h_0(\qsq) H_0(\qsq)$.  The form factors are normalized such that $\qsq{} \Hszer \rightarrow 1$
as $\qsq{} \rightarrow 0$.
 \label{qsqhsq}}
 \end{center}
\end{figure}

Figure \ref{qsqhsq} gives a different insight into the helicity basis form factors by
plotting the intensity contributions of each of the form factor products. This is the
form factor product multiplied by \qsq{}. 
Since \qsq{}\Hszer{} dominates, we normalized form factors such that $\qsq{}\Hszer{} = 1$ at \qsq{} = 0 but use the
same scale factor for the other three form factors.
As expected, both \qsq{} \Hspls{} and \qsq{}\Hsmin{} rise from zero with increasing \qsq{} and they both appear to approach \qsq{}\Hszer{} at  \qsq{}$_{max}$ -- although \qsq{} \Hspls{} seems slightly lower than
\qsq{} \Hszer{} at \qsq{}$_{max}$.

\begin{figure}[tbph!]
 \begin{center}
  \includegraphics[width=3.in]{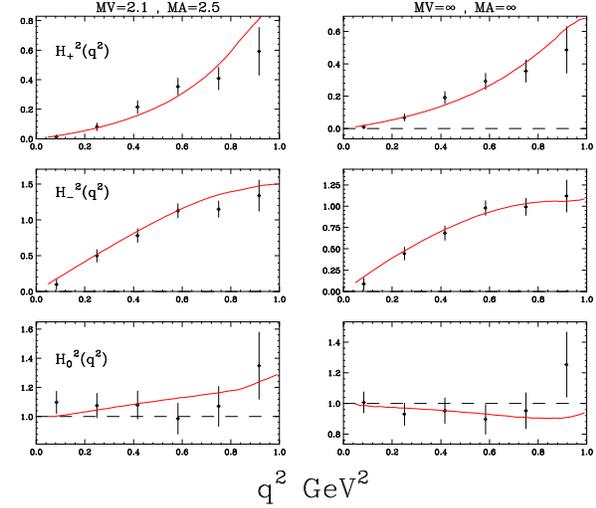}
  \caption{Non-parametric form factor products obtained for data (multiplied by \qsq{})  
The solid curves are based on the $s$-wave model and measurements described in Reference~\cite{formfactor}. 
The reconstructed form factor products are the points with error bars. The three plots on the right are the usual model with the 
spectroscopic pole masses; while the three plots on the right are run with the axial and vector
pole masses taken to infinity. \label{infpole}}
 \end{center}
\end{figure}

What can we learn about the pole masses?  Unfortunately Fig. \ref{infpole} shows that the present
data is insufficient to learn anything useful about the pole masses.  On the 
left of Figure \ref{infpole}, the helicity form factors are compared to 
a model generated with the FOCUS form factor ratios\cite{formfactor} and the standard pole masses
of 2.1 GeV for the vector pole and 2.5 GeV for the two axial poles. On the right side
of Fig. \ref{infpole}, the form factors are compared to a model where the pole masses
are set to infinity meaning that the axial and vector form factors are constant.  Both models fit the data equally well.

\begin{figure}[tbph!]
 \begin{center}
  \includegraphics[width=2.7in]{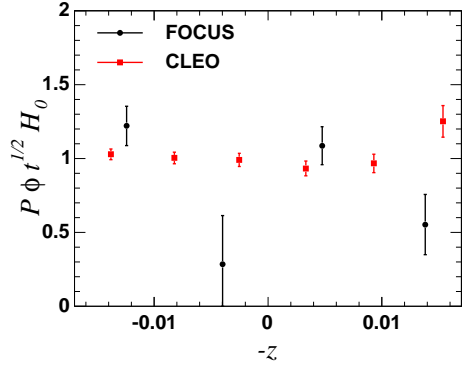}
  \caption{Transformation of \Hzer{} into $H_0(z)$ by R.J. Hill\cite{Hill2}.  Here $t \equiv \qsq$ and $P$ and $\phi$ are functions of \qsq{} designed
to remove the simple poles.
The FOCUS data is from Reference
\cite{focus-helicity} and the CLEO data is from Reference \cite{cleo-ff}.
\label{hillH0}} \end{center} \end{figure}

The data of Fig. \ref{infpole} is consistent with the spectroscopic pole dominance albeit with essentially no 
sensitivity to the pole masses.  Fig. \ref{hillH0} shows that it is also consistent with the expected behavior
under a Hill transformation, illustrated earlier in Fig. \ref{hillTrans}. 
Fig. \ref{hillH0} shows the result of transforming from \qsq{} to $z$ according to the Hill prescription\cite{Hill2}.
Over the very narrow $-z$ range accessible for \kpilndk{} , it is not surprising that that the transformed form factor is
essentially constant.

It is interesting to note that the FOCUS analysis was based on a sample of 11400 \kpimndk{} events, while the CLEO
analysis was based on a sample of only 2470 \kpiendk{} events. The error bars in Fig. \ref{hillH0} for FOCUS data are much larger than those for the much smaller CLEO data set and only four FOCUS \qsq{} bins are reported on. This is because of the much poorer \qsq{} resolution in fixed target semileptonic decay compared to the order-of-magnitude better \qsq{} resolution obtainable for semileptonic analyses in charm threshold data from $e^+ e^-$ colliders where the neutrino can be reconstructed using energy-momentum balance. This is especially relevant for \kpilndk{}
since the 1 GeV$^2$ \qsq{} range for \kpilndk{} is a factor of two smaller than that in \klndk{}. Error inflation due to 
deconvolution grows dramatically once the bin-to-bin separation, $\Delta \qsq{}$, approaches the r.m.s. resolution, $\sigma(\qsq)$, which was
typically 0.18 GeV$^2$ in the four bins reported on by FOCUS\cite{focus-helicity}.

\begin{figure}[tbph!]
 \begin{center}
  \includegraphics[width=2.3in]{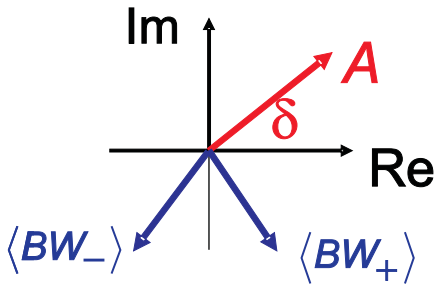}
  \caption{Illustration of s-wave phase
\label{swave-cartoon}} \end{center} \end{figure}

What can we learn about the phase of the s-wave contribution?  Recall in 
Figure \ref{asym} the asymmetry created by the interference between the
s-wave and \krzlndk{} only appeared below the \krzb{} pole in FOCUS data and thus
the s-wave phase was such that it was orthogonal with the $\mkpi > m(\krzb)$ half
of the  Breit-Wigner amplitude or $\langle BW_+ \rangle$. Since the asymmetry is ``negative" according to the 
convention of Eq. (\ref{KSE}), in that favors the backward over the forward \costhv{}
direction, it must be anti-collinear to $\langle BW_- \rangle$ as well. Hence it must have roughly the phase of 40$^0$ as illustrated by Fig. \ref{swave-cartoon}. FOCUS\cite{formfactor} measured the s-wave phase to be $\delta = (39 \pm 4 \pm 3)^0$.

\begin{figure}[tbph!]
 \begin{center}
  \includegraphics[width=3.in]{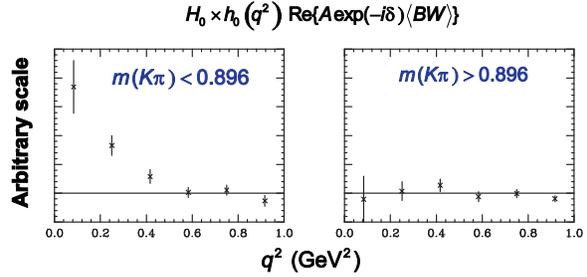}
  \caption{The s-wave interference term for events below the \krzb pole (left) and above the pole (right).
The interference term depends on the s-wave phase relative to the phase average phase of each 
half of the Breit-Wigner.  All of the \costhv{} interference observed by FOCUS was also below the 
\krzb pole as shown in Fig. \ref{asym}
 \label{split}}
 \end{center}
\end{figure}

As Figure \ref{split} shows, the same thing happens in CLEO data.
The effective \Hint{} disappears above the \krzb pole and is very strong
below the pole.  The amplitude $A$ of the s-wave piece is arbitrary since
using interference we can only observe the product $A~H_0 (\qsq{})~h_0 (\qsq{})$.  This
means any change in $A$ scale can be compensated by a change of scale in $h_0 (\qsq{})$.
The fact that the \Hint{} data was a tolerable match (at least in the low \qsq{} region) to the FOCUS curve in Figure \ref{hsq} 
does imply, however, that the s-wave amplitude observed in CLEO is consistent with that of FOCUS.  A more formal
fit of the s-wave parameters in CLEO data is in progress.

\begin{figure}[tbph!]
 \begin{center}
  \includegraphics[width=3.in]{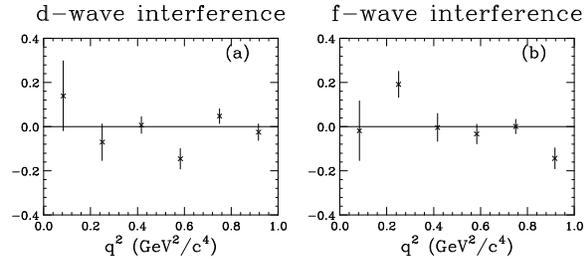}
  \caption{Search for (a) $d$-wave and (b) $f$-wave interference effects as described in the text.
 \label{fdwave}}
 \end{center}
\end{figure}

Finally, is there evidence for higher $K^- \pi^+$ angular momentum amplitudes in \kpilndk{}? 
We searched for possible additional interference terms such as a (zero helicity) d-wave contribution:
$4\,\sinthlsq (3\,\cos^2 \thv - 1)\,H_0(q^2)\,h^{(d)}_0(q^2)\,
{\mathop{\mathrm{Re}}\nolimits}\{\mathrm{A}e^{-i\delta} \bw\}$  or
an f-wave contribution: 
$4\,\sinthlsq \cos \thv (5\,\cos^2 \thv - 3)\,H_0(q^2)\,h^{(f)}_0(q^2)\,
{\mathop{\mathrm{Re}}\nolimits}\{Ae^{-i\delta} \bw \}$. 
As shown in Figure \ref{fdwave}, there is no evidence for such additional contributions which should diverge
as $1/\qsq{}$ at low \qsq{}.

\section{Future Directions}
It will be interesting to pursue the non-parametric \kpilndk{} analysis with more data.  One motivation is 
will be to further study the $h_0(\qsq)$ form factor which appears to be somewhat different than
$H_0(\qsq)$. It would also be interesting to pursue tighter limits on possible d-wave and f-wave non-resonant contributions
to \kpilndk{} and make more stringent tests of SPD.
CLEO is slated to increase their luminosity at the $\psi(3770)$ from the 280 pb$^{-1}$ reported here to 750 pb$^{-1}$.
In addition Surik Mehrabyan and I, are studying \kpimndk{} as well as \kpiendk{} in CLEO data. This is
a somewhat challenging project since the CLEO muon detector was designed for higher energy B-meson running and the muons
from charm semileptonic decay tend to range out before being identified. Hence special care must be exercised to 
reduce backgrounds. Besides increasing our statistics, the \kpimndk{} should allow us to make the first measurements
of the $H_T(\qsq)$ form factor which is suppressed by a factor of $m^2_\ell/\qsq$. Since this is a zero helicity factor,  it can interfere
with \Hzer{} and hence two new projectors will be required: one for the $H_T^2(\qsq)$ term and one for $H_0(\qsq) \times H_T(\qsq)$ interference. At present the prognosis for making these measurements looks good.

\mysection{Summary}

Progress in understanding \veclndk{} decays was reviewed.  These have
historically been analyzed under the assumption of spectroscopic pole dominance (SPD).  
A recent result from BaBar was reviewed that used SPD to show that the form factors for \philndk{} are
consistent with those from \kpilndk{} as expected from SU(3) symmetry.
Experiments
have obtained consistent results with the SPD assumption, but as of yet there have been
no incisive tests of spectroscopic pole dominance.   We concluded by describing a first non-parametric look at the \kpilndk{} form factors.  Although the results were very consistent
with the traditional pole dominance fits, the data was not precise enough to incisively
measure \qsq{} dependence of the axial and vector form factors and thus test SPD.  This
preliminary analysis did confirm the existence of an $s$-wave effect first observed by FOCUS \cite{swave},
but was unable to obtain evidence for $d$ and $f$-waves.

\end{document}